\documentclass[twocolumn]{webofc}

\usepackage[varg]{txfonts}   
\usepackage{hyperref}
\usepackage{url}
\hypersetup{colorlinks=true,citecolor=blue,urlcolor=blue,linkcolor=blue}
\begin{document}
%
\title{Timing Challenges in Transmission Silicon Detectors for Light Ions}
%
%

\author{\firstname{Diego} \lastname{Tarr\'io}\inst{1}\fnsep\thanks{\email{diego.tarrio@physics.uu.se}}
\and
        \firstname{Friedrich Karl} \lastname{Moraht}\inst{1} \and
        \firstname{Lucas} \lastname{De Arruda}\inst{1,2,3} \and
        \firstname{Stephan} \lastname{Pomp}\inst{1} \and
        \firstname{Xavier} \lastname{Ledoux}\inst{2} \and
        \firstname{Diego} \lastname{Ramos}\inst{2}
}

\institute{
Department of Physics and Astronomy, Uppsala University, Box 524, 75120 Uppsala, Sweden
\and
Grand Accélérateur National d’Ions Lourds, CEA/DRF-CNRS/IN2P3, B.P. 55027, 14076, Caen, France 
\and
Université de Caen-Normandie, Normandie Université, 14000 Caen, France
          }

\abstract{
Transmission silicon detectors are widely used in $\Delta E$--$E$ telescope configurations for particle identification and timing in nuclear physics experiments. Variations in signal rise time can introduce time-walk effects that degrade time-of-flight measurements, as recently observed in light-ion production experiments at the Neutrons For Science (NFS) facility at GANIL using the Medley setup. This work investigates the influence of particle energy and detector non-uniformity on signal rise time.
}

\maketitle

\section{Introduction}\label{sec:intro}
Transmission silicon detectors are widely used in $\Delta E$--$E$ and $\Delta E$--$\Delta E$--$E$ telescope configurations for light-ion identification in nuclear physics experiments. In such systems, silicon detectors can also provide timing information for time-of-flight (TOF) measurements.

A representative example is the Medley experimental setup, developed at Uppsala University and now installed at the Neutrons For Science (NFS) facility at GANIL (Caen, France). Medley is used to measure cross sections for light-ion production in neutron-induced reactions using a set of $\Delta E$--$\Delta E$--$E$ telescopes covering different emission angles \cite{DeArruda2026}. Timing signals from the second silicon detector in each telescope allow the incident neutron energy to be determined by TOF.


During the first experimental campaigns with Medley at NFS, deviations from the expected timing behaviour were observed, indicating systematic effects in the detector time response. Although these effects can be partially corrected through empirical methods~\cite{DeArruda2025}, further investigation is needed to mitigate their impact in future experiments.

The present work combines in-beam observations from Medley at NFS with off-beam detector-characterisation measurements. The first part examines timing effects associated with particle energy, particularly around the punch-through threshold. The second part investigates whether detector response non-uniformities can also contribute to rise-time variations. Together, these studies aim to clarify the origin of the timing distortions observed in transmission silicon detectors.

\section{Experimental methods}\label{sec:methods}
\subsection{Medley at NFS: in-beam measurements}\label{sec:medley}

\begin{figure}[h!]
\centering
\includegraphics[width=1.\columnwidth,clip]{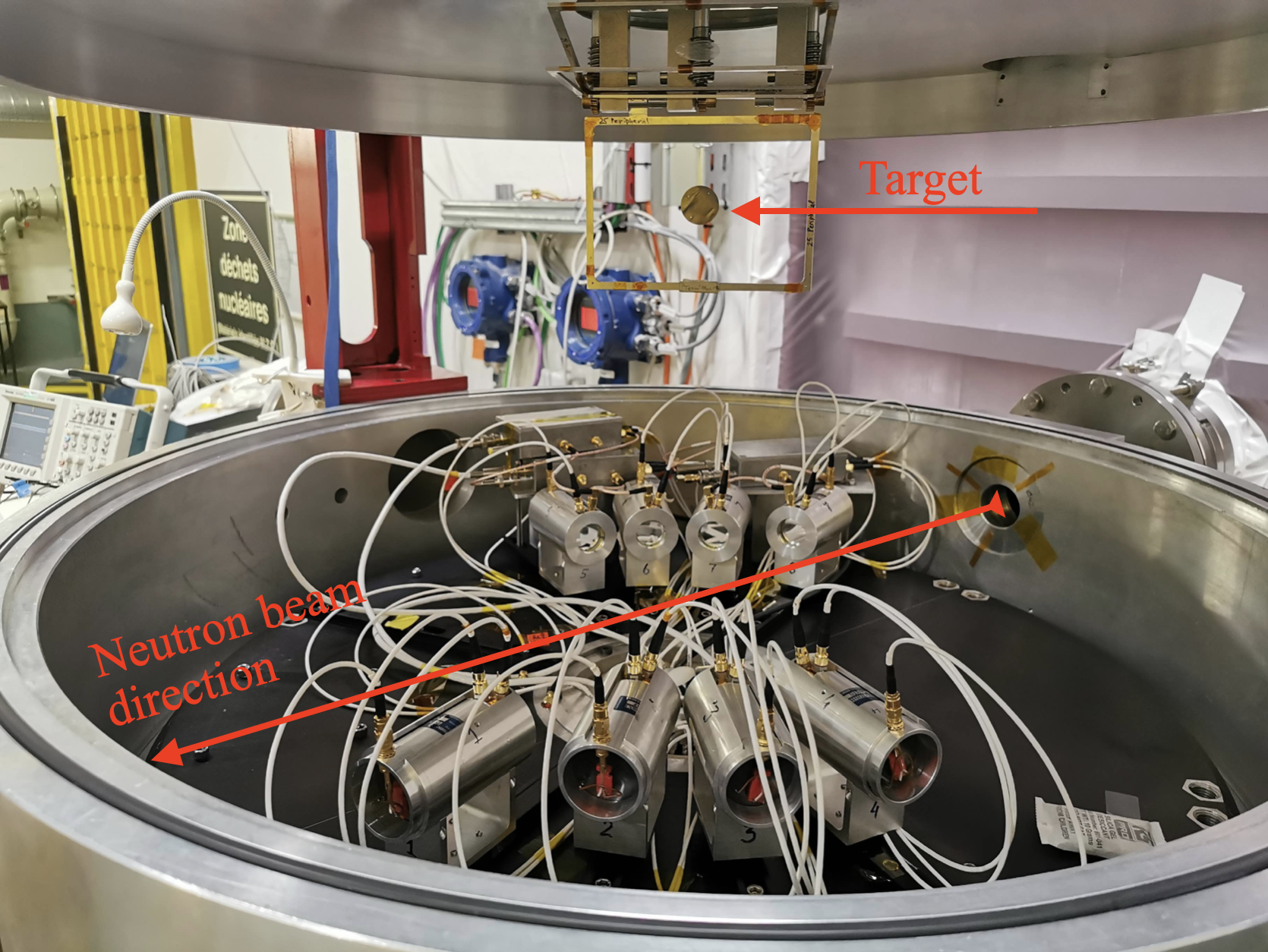}
\caption{Inner view of the open Medley chamber. Eight telescopes point to the central position of the chamber where the target is placed when the lid is closed. The direction of the neutron beam is indicated.}
\label{fig:Medley}
\end{figure}

The NFS facility provides a neutron beam with a continuous energy spectrum extending from approximately \mbox{1 to 40 MeV}~\cite{Ledoux2021}. Installed at NFS, the Medley setup is designed to measure light-ion production cross sections in neutron-induced reactions over this energy range~\cite{DeArruda2026}. Medley consists of eight detector telescopes housed in a vacuum chamber, each comprising two silicon transmission detectors followed by a CsI(Tl) crystal in a $\Delta E$--$\Delta E$--$E$ configuration (see Fig.~\ref{fig:Medley}).

The telescopes are arranged around the target at 20$^\circ$ intervals, and the entire assembly can be rotated to cover different emission angles. In telescopes 1 to 4, which typically cover the forward angles, the first and second silicon detectors are about 50~$\mu$m and 1000~$\mu$m thick, respectively, whereas in telescopes 5 to 8, which usually cover the backward angles, they are 20~$\mu$m and 500~$\mu$m thick. All CsI detectors are 5~cm long, sufficient to stop the highest-energy protons.

Signals are processed using charge-sensitive preamplifiers and digitized with the Numexo2 system~\cite{Houarner2025}. Timing information is extracted from the signal of the second silicon detector in each telescope using a constant-fraction discriminator (CFD)-based digital algorithm implemented in Numexo2. For neutron-energy determination through time-of-flight measurements, this signal defines the start time, while the radio-frequency signal from the subsequent accelerator pulse provides the stop time. Only pulse-height and timing information are stored for offline analysis.

\subsection{Off-beam detector-response mapping}\label{sec:mapping_setup}

In addition to in-beam observations, a dedicated setup was developed to investigate detector-response non-uniformities independently of punch-through effects. The measurements were performed using spare detectors identical to those installed in the Medley setup at NFS. Similar behaviour is therefore expected, although variations between individual detectors cannot be excluded.

Silicon detectors were irradiated with a mixed $\alpha$ source ($^{239}$Pu, $^{241}$Am, $^{244}$Cm), providing particles with discrete energies that fully stop within the detector~\cite{Moraht2025}. This allows the separation of position-dependent effects from those related to punch-through.

The source was mounted, with a 2-mm diameter collimator, in a motorized positioning system to scan each detector across its surface. The measurements were carried out in a vacuum chamber at a typical pressure of $10^{-2}$~mbar. The detector was connected to a charge-sensitive preamplifier (of the same type as used at NFS) and full waveforms were recorded using a 1-GHz digitizer for offline analysis. The details of this setup and its performance have been extensively described in Ref.~\cite{Moraht2025}.

The stored waveforms have been analysed using ROOT-based software to extract the pulse height and the rise time, defined as the time required for the signal to increase from 10\% to 90\% of its maximum amplitude. Approximately $10^4$ events were recorded at each position for each of the investigated detectors.

\section{Energy-dependent timing effects in Medley}\label{sec:energy}
Neutron–proton elastic scattering provides an ideal time-of-flight reference due to the well-known linear relation between the scattered proton energy E$_p$ at angle $\theta$ and the incoming neutron energy E$_n$:
\begin{equation}
    E_p(\theta)= \frac{4 m_p m_n}{(m_p+m_n)^2}E_n \cos^2 \theta\approx  E_n \cos^2 \theta
\label{eq:scattering}
\end{equation}
where $m_p$ and $m_n$ are the proton and neutron masses, respectively. The approximation $m_p\approx m_n$ has been used.

During the experimental campaigns, a CH$_{2}$ target was used to measure the neutron spectral flux. Scattered protons are detected and identified in the telescope at 20$^\circ$, where their kinetic energy is measured by the three detectors of the telescope; the neutron energy is determined using the time-of-flight (TOF) technique as described in Section~\ref{sec:medley}. The TOF of the proton between the CH$_{2}$ target and the detector is also taken into account~\cite{DeArruda2025}.

\begin{figure}[h]
\centering
\includegraphics[width=\columnwidth,clip]{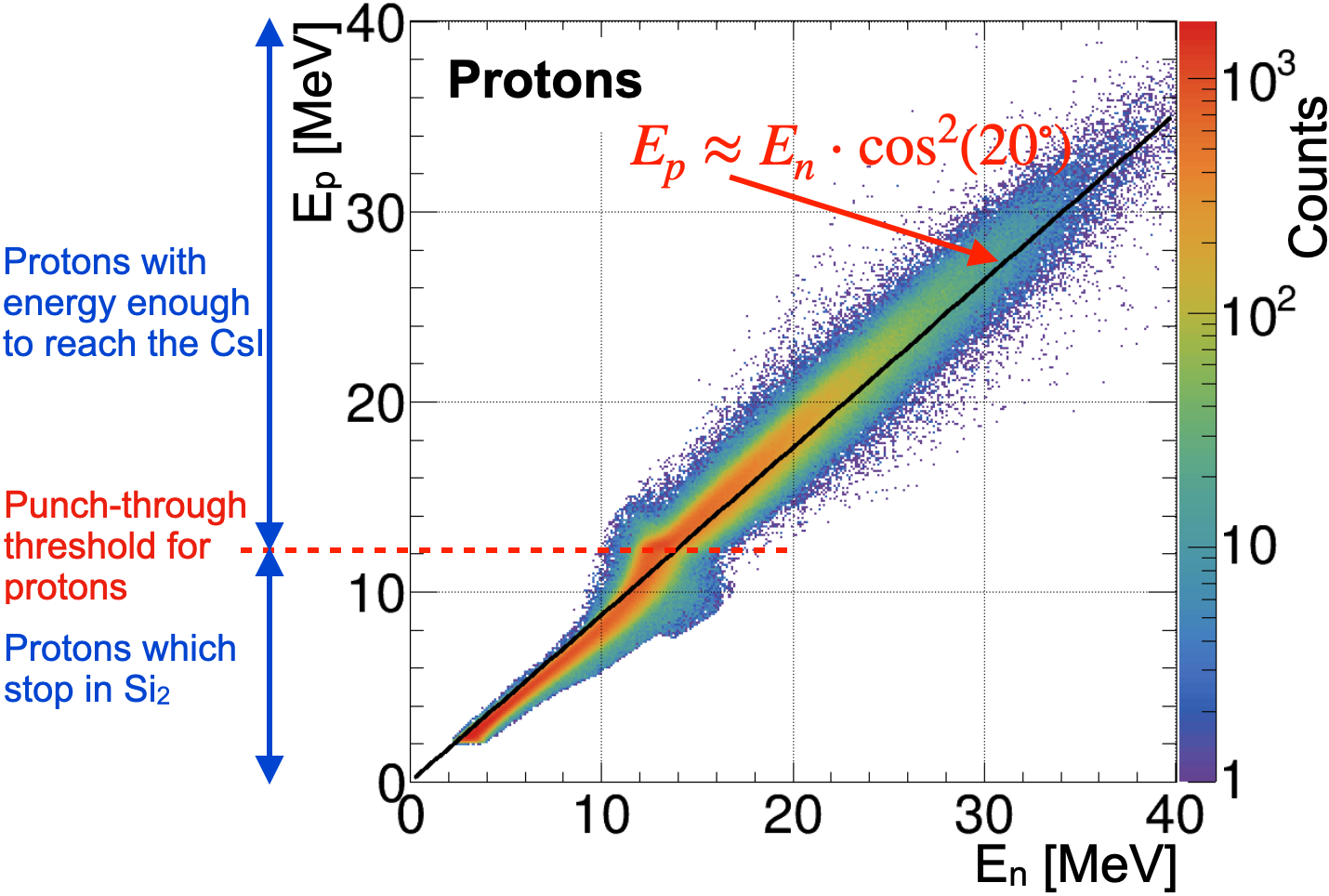}
\caption{Experimental correlation between proton and neutron energies in neutron--proton elastic scattering observed with Medley at NFS in the telescope placed at $20^{\circ}$. Deviations from the expected linear relation from Eq.~\ref{eq:scattering} (solid line) are observed around the punch-through threshold. Based on data from Ref.~\cite{DeArruda2025}.}
\label{fig:EpEn}
\end{figure}

As shown in Fig.~\ref{fig:EpEn}, the relationship between the scattered proton energy and the neutron energy deviates from the expected linear behaviour. The strongest deviation appears for protons of about 12.3~MeV, corresponding to the minimum energy at which protons punch through the second silicon detector. Similar deviations from linearity are observed for deuterons and tritons around their respective punch-through thresholds (16.6 and 19.5~MeV, respectively). This behaviour is observed consistently across all telescopes, despite their different detector thicknesses and, consequently, different punch-through thresholds.

The deviation around the punch-through threshold suggests that the pulse shape depends on the charge deposition depth in the detector, affecting the timing response.

\section{Physical origin of rise-time variations in transmission silicon detectors}\label{sec:signal}

To interpret these deviations, consider signal formation in a silicon detector. When a charged particle enters the detector, it loses energy through ionisation, creating electron--hole pairs along its path. Under the electric field, electrons drift to the anode and holes to the cathode. The signal time evolution depends on the spatial distribution and drift paths of the charge carriers, and thus on the depth at which the charge is generated~\cite{England1995}.

\begin{figure}[h]
\centering
\includegraphics[width=\columnwidth,clip]{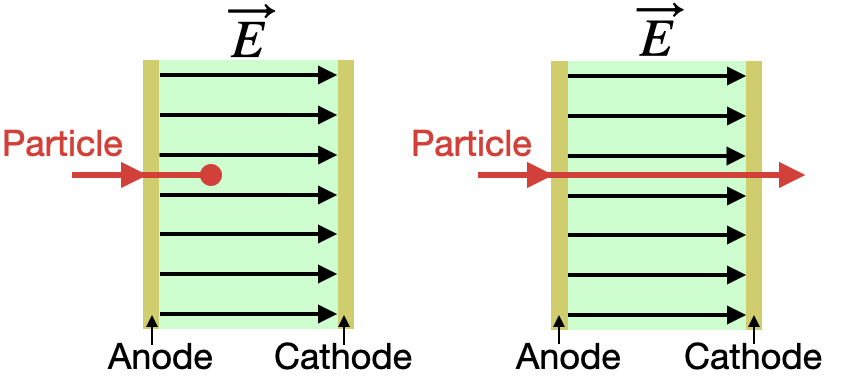}
\caption{Schematic illustration of a particle impinging on a silicon detector from the anode side. \textit{Left:} The particle stops within the active volume; \textit{Right:} The particle punches through the detector, producing ionisation throughout the full detector thickness.}
\label{fig:SiParticles}
\end{figure}

A key difference arises depending on whether the particle stops inside the detector or punches through it (see Fig.~\ref{fig:SiParticles}). For a particle stopping inside the detector, charge is generated only up to the stopping point. The carriers released along the trajectory drift towards the electrodes; the drift of electrons towards the anode (in our case, the entrance window) determines the rise time of the pulse. Higher-energy particles penetrate deeper, increasing the contribution from carriers created farther from the anode and resulting in longer signals with slower rise times. When the particle energy is sufficient to punch through the detector, ionisation occurs throughout the full detector thickness. Beyond this point, further increases in energy have little effect, and the rise time becomes nearly independent of the particle energy.

Since the pulse rise time depends on the charge deposition depth and on charge carrier drift, it is also expected to be sensitive to spatial variations in the detector response, such as local electric-field non-uniformities or partially depleted regions. This motivates the dedicated detector-mapping measurements presented in the next section.

\section{Spatial response and detector non-uniformities}\label{sec:nonuniformity}

\begin{figure*}
\centering
\includegraphics[width=\textwidth,clip]{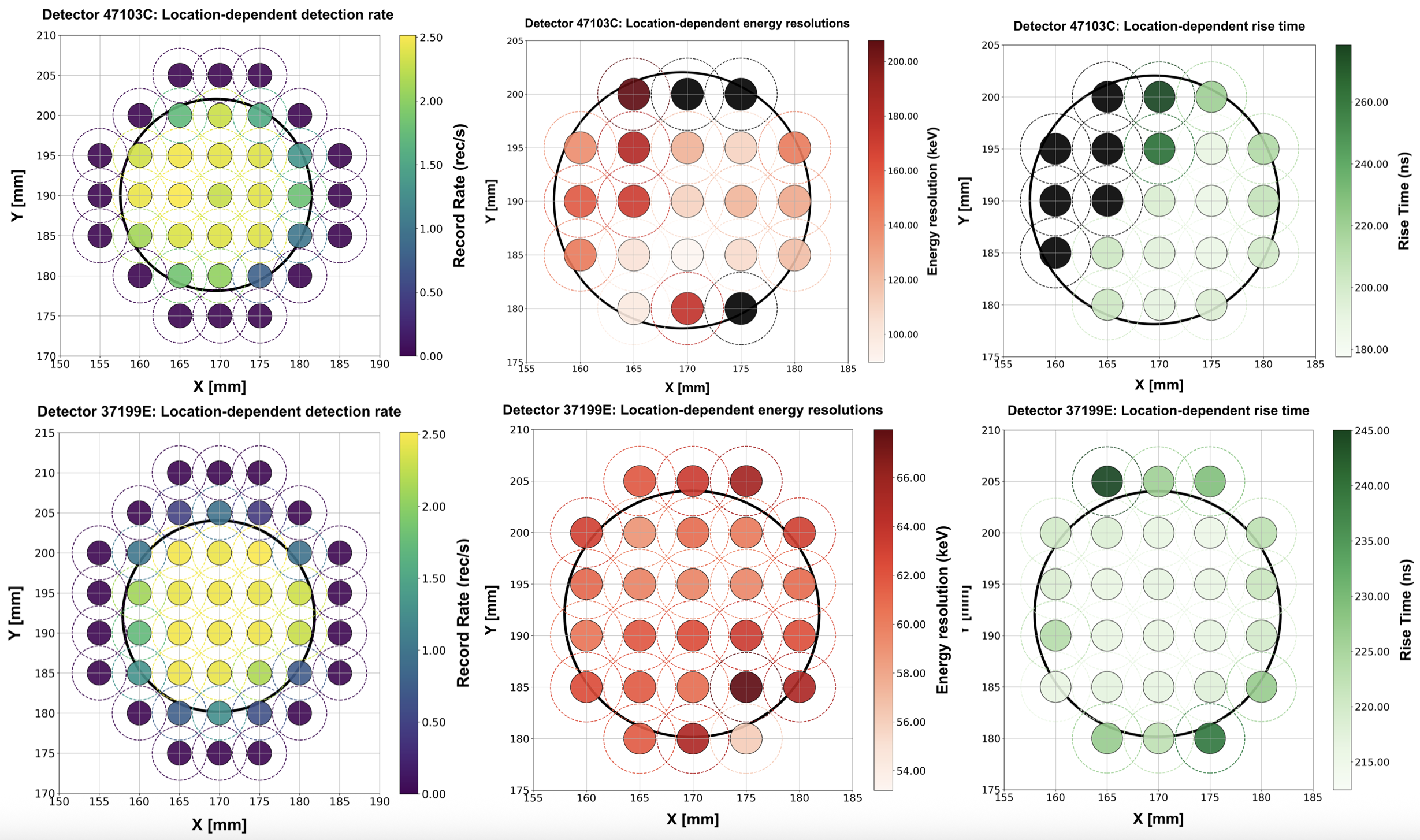}
\caption{Spatial mappings for two silicon detectors, 1000~\textmu m thick (top) and 400~\textmu m thick (bottom). Black circles indicate the detector size, filled circles the measurement positions, and dashed circles the irradiated area. Left: counting rate; centre: energy resolution; right: rise time. Adapted from Ref.~\cite{Moraht2025}.}
\label{fig:Mapping}       
\end{figure*}

Spatial mappings for two silicon detectors of different thicknesses are shown in Fig.~\ref{fig:Mapping}. The top row corresponds to a 1000~\textmu m thick detector, and the bottom row to a 400~\textmu m thick detector. In each row, the counting rate (left), energy resolution (centre), and rise time (right) are shown for a series of positions across the detector front surface. The large black circle in each panel indicates the detector size (24~mm diameter). Each filled circle represents one measurement point, with a diameter corresponding to the 2~mm collimator aperture.

For the 1000~\textmu m thick detector, the rise time (Fig.~\ref{fig:Mapping}, top right) ranges from approximately 180~ns to 1.4~$\mu$s and shows a strong position dependence, with values increasing towards the left side of the detector. A similar spatial pattern is observed in the energy resolution (top centre), which degrades from 89~keV (1.6\%) up to 320~keV (5.8\%), suggesting a common underlying origin for both behaviours. In contrast, the 400~\textmu m thick detector (bottom row) exhibits a more uniform rise time (from 216 to 245~ns) and energy resolution (from 56.2 to 68.0~keV, i.e. from 1.0\% to 1.2\%). For both detectors, the counting rate remains uniform across the surface.

These measurements demonstrate that substantial position-dependent variations in rise time can occur even for alpha particles that are fully stopped within the detector. Energy-dependent effects related to punch-through therefore cannot be the sole source of timing variations in transmission silicon detectors.

\section{Discussion}\label{sec:discussion}
The in-beam measurements with Medley show that the timing response is strongly affected by the particle energy, in particular around the punch-through threshold of the second silicon detector. The behaviour observed in Fig.~\ref{fig:EpEn} is consistent with the signal-formation model discussed in Section~\ref{sec:signal}, in which the rise time depends on the depth at which charge carriers are generated. In this respect, energy-dependent pulse-shape variations provide a natural explanation for the deviations observed in the proton--neutron energy correlation.

The off-beam measurements, however, show that energy-dependent effects alone cannot account for all rise-time variations. The spread observed for alpha particles of well-defined discrete energies demonstrates that detector non-uniformities can also play a major role in the timing response. In the 1000~\textmu m thick detector, the spatial pattern of the rise time is similar to that observed for the energy resolution, suggesting a common underlying origin. Possible causes include spatial variations in the resistivity of the detector material, local non-uniformities in the electric field, partially depleted regions, or variations in dead-layer thickness. The present measurements do not allow distinguishing between them.

The comparison with the 400~\textmu m thick detector suggests that the thicker detector is more strongly affected by inhomogeneities. More generally, the results are consistent with the possibility that such effects become more significant in thicker silicon detectors, although a broader study involving a larger number of detectors would be required to establish this conclusion.

These results show that the timing response in Medley depends not only on particle energy but also on the interaction position in the detector. The dedicated off-beam measurements were performed on spare detectors of the same type as those used in the experiment and therefore support the interpretation of detector non-uniformity as an additional contributor to time walk, even though a direct detector-by-detector correlation with the in-beam measurements has not been done.

In the Medley experiments, the timing extraction is performed using a digital implementation of a constant-fraction discriminator (CFD) integrated into the Numexo2 acquisition system, which determines the signal time at a fixed fraction of its amplitude. While this method is usually effective in compensating for time walk associated with amplitude variations, it relies on the assumption that the signal shapes are sufficiently similar~\cite{Kno10}. Large variations in pulse shape and rise time, such as those observed here, can therefore limit the effectiveness of CFD techniques in correcting for time walk.


\section{Conclusions and outlook}\label{sec:conclusions}
Timing challenges in transmission silicon detectors have been investigated through a combination of in-beam measurements with Medley at NFS and dedicated off-beam detector characterisation studies.

The in-beam data show that particles with energies below and above the punch-through threshold produce signals with different rise times, thereby affecting time-of-flight measurements. In addition, off-beam alpha-source measurements reveal significant spatial variations in detector rise time, with a pattern similar to that observed for the energy resolution. These observations demonstrate that both particle range and detector non-uniformities strongly influence the signal shape and, consequently, the timing response, and must be accounted for in future detector operation and analysis.

Although empirical correction factors have proven useful in the analysis of previous Medley campaigns at NFS, future experiments would benefit from alternative approaches based on pulse-shape analysis to account more explicitly for rise-time variations.

\section{Acknowledgments}\label{sec:acknowledgements}
Part of this work has been carried out within the framework of the EUROfusion Consortium, funded by the European Union via the Euratom Research and Training Programme (Grant Agreement No 101052200 — EUROfusion). This work has also been supported by the Swedish Radiation Authority (project number SSM2024-1205-17), by the MÅSTE project (P2023-01281) funded by the Swedish Energy Agency, and by the Swedish Center for Nuclear Technology (SKC).

\end{document}